# CONDITIONS FOR STIMULATED EMISSION IN ANOMALOUS GRAVITY-SUPERCONDUCTORS INTERACTIONS

*Giovanni Modanese[1] and Timo Junker[2]*
[2] Göde Wissenschaftsstiftung
Am Heerbach 5, 63857 Waldaschaff, Germany
[1] University of Bolzano – Logistics and Production
Engineering; Via Sernesi 1, 39100 Bolzano, Italy

## ABSTRACT

Several authors have studied the generation of gravitational fields by condensed-matter systems in non-extreme density conditions (i.e., conditions not like those of collapsed stars, but such to be possibly obtained in a laboratory). General Relativity and lowest-order perturbative Quantum Gravity predict in this case an extremely small emission rate, so these phenomena can become relevant only if some strong quantum effect occurs. Quantum aspects of gravity are still poorly understood. It is believed that they could play a role in systems which exhibit macroscopic quantum coherence, like superconductors and superfluids, leading to an "anomalous" coupling between matter and field. We mention here recent work in this field by Woods, Chiao, Becker, Agop et al., Ummarino, Kiefer and Weber. Many of these theoretical works were stimulated by the experimental claims of Podkletnov. His results have not yet been confirmed, but the published replication attempts have admittedly been incomplete. Recently, Tajmar claimed to have detected a gravitomagnetic field generated by a spinning superconductor. Chiao also made some attempts at the construction of a gravity/e.m. transducer based on quantum effects. In our previous theoretical work, we sought an interpretation of the anomalous emission reported by Podkletnov as a consequence of the local modification of the vacuum energy density in the superconductor. We hypothesed that the vacuum energy density term could interfere with a set of strong gravitational fluctuations called "dipolar fluctuations". In this chapter we improve our earlier model and also present new results concerning anomalous stimulated gravitational emission in a layered superconductor like YBCO. We model the superconductor as an array of intrinsic Josephson junctions. The superconducting parameters are defined by our preliminary measurements with melt-textured samples. Coherent e.m. emission by synchronized Josephson junctions arrays was first reported by Barbara et al. in 1999. We write explicitly and solve numerically the Josephson equations which give the normal and super components of the total current in the superconductor, and derive from this the total available power $P=IV$. Then, assuming that the coefficients $A$ and $B$ for spontaneous and stimulated gravitational emission are known, we apply to this case the Frantz-Nodvik equation for a laser amplifier. The equation is suitably modified in order to allow for a "continuous pumping" given by an oscillating transport current. The conclusions are relevant for the evaluation of gravitational emission from superconductors. We find that even if the $A$ and $B$ coefficients are anomalously large (possibly because of the Quantum Gravity effects mentioned above), the conditions for stimulated emission are quite strict and the emission rate strongly limited by the *IV* value, for reasons intrinsic to the nature of the superconductor.

## 1. INTRODUCTION

Over the last decades several authors, mainly from the General Relativity community, were intrigued by the idea that the interaction between gravity and superconductors might be somehow peculiar. The simplest proposals were about using superconductors as sensitive field detectors. Some also speculated, however, that superconductors could act as effective emitting antennas of gravitational waves. This is clearly outside the orthodoxy of General Relativity, which "weighs" any gravitational source only with regard to its energy-momentum, independently from its microscopic structure or composition.

But if gravitation has to be eventually reconciled with quantum mechanics, the macroscopic quantum character of superconductors might actually matter. In a recent authoritative review on the "Interaction of gravity with mesoscopic systems" [1], Kiefer and Weber recall that the interaction of gravitational fields with quantum fluids has been extensively studied. They mention work published by De Witt in 1966, Papini in 1967, Anandan and Chiao in 1981-84, Peng and Torr in 1990-91. Then they focus on the ideas which describe generation and detection of gravitational waves via the use of quantum fluids. They investigate the arguments suggesting that quantum fluids should be better interaction partners of gravitational waves than classical materials, and discuss proposed coupling schemes, including those by De Matos and Tajmar [2] and by Chiao et al. [3].

The "HFGW conferences" (High Frequency Gravitational Waves) held in 2003 and 2007 collected numerous conservative and speculative works. Conservative works typically involve General Relativity estimates of (very low) gravitational waves power emitted by laboratory devices with high-frequency vibrations. Speculative works (for instance, [4]) hypotesized that the graviton emission amplitude is somehow amplified by quantum properties of matter. Woods [5] discusses impedance mismatch at superconductor-air interfaces in the propagation of HFGWs. In the case of type-II superconductors with variable internal magnetization, he shows that this amounts to a sizeable interaction between the gravitational wave and the magnetic field ("enhanced Gertsenshtein effect"). He argues that this may be exploited for the design of a novel type of lens for HFGWs, using a magnetic field to adjust the focal length.

Agop et al. [6] write equations for a generalized Meissner effect, which take into account the gravito-magnetic and gravito-electric fields in the Maxwell-Einstein approximation. They find a very large "gravitational screening length", in accordance with previous authors, and yet their screening equation also involves the short length $\lambda_e \approx 10^{-8}$ m. Ummarino [7] calculates the possible alteration of the gravitational field in a superconductor using the time-dependent Ginzburg-Landau equations and compares the behaviour of a high-$T_c$ superconductor like YBCO with a classical low-$T_c$ superconductor like Pb.

Many of these theoretical works were stimulated by the experimental claims of Podkletnov [8,9]. His results have not yet been confirmed, but the published replication attempts have admittedly been incomplete [10].

One of the problems of current models of gravity-superconductors interaction is the over-simplified representation of the superconductors. Type-II superconductors and anisotropic ceramic superconductors have a complex microscopic structure, far from the ideal fluid model suitable for type-I superconductors with long-range coherent wave functions. In this work we shall model a melt-textured YBCO emitter according to the intrinsic Josephson junctions picture which has been firmly established starting from the '90.

In general terms, we believe that gravitational emission from superconductors is limited by three main factors.

1. The fundamental coupling. A quantum mechanism is needed, which escapes the severe limitations of the standard General Relativity coupling. We have proposed earlier such a dynamical mechanism, based on the vacuum fluctuations of Quantum Gravity. Here we recall it briefly and add some new remarks. This mechanism is only able to generate a strong virtual radiation, ie off-shell gravitons with $\lambda f \ll c$ and finite propagation range (but also with spin 0 and 1 components).
2. Energetic efficiency. This is a key limitation even in electromagnetic Josephson emission from superconductors. The maximum available power $P=IV$ is usually small, due to very small voltage drops in good superconductors. In addition, any current injected from the outside causes considerable dissipation at the superconducting-normal contacts.
3. Stimulated emission. It has been found experimentally [26] that electromagnetic emission from Josephson junctions arrays can be amplified by stimulated emission. In the absence of any resonant cavity suitable for gravitational radiation, this can only occur in a single-pass mode, like in an optical or maser amplifier governed by the Frantz-Nodvik rate equation. The non-standard dispersion relation $\lambda f \ll c$ implies that real stimulated electromagnetic emission does not compete with virtual graviton emission.

We shall address the Points (1), (2) and (3) above in Sec. 2, 3, 4, respectively. The techniques employed vary. Sec. 2 mainly involves Quantum Gravity considerations and a model of gravitational vacuum fluctuations. We recall previous work and prove, or a least justify, a chain of equivalences: (i) the presence of a superconductor amounts to a local variation in the vacuum energy density $\Lambda$; (ii) a time-variable $\Lambda$ is equivalent to an oscillating virtual mass $M_{\Lambda,\text{eff}}$; (iii) this means in turn that in a Josephson junction under high-frequency current, sizeable coefficients of spontaneous and stimulated graviton emission can be defined. The weakest link in this chain is the proof of the "amplification" $M_{\Lambda,\text{eff}} \gg M_\Lambda = \Lambda V/G$, where $V$ is the spatial volume of the region where $\Lambda$ is present, and the vacuum energy density $\Lambda/G$ associated with a superconductor is typically $10^6$-$10^8$ J/m$^3$.

Sec. 3 employs notions and techniques from the theory and experimental practice of ceramic superconductors and Josephson junctions. On the base of data from our preliminary measurements, an YBCO emitter is analysed as a series of intrinsic Josephson junctions, whose behaviour is numerically simulated within the RSJ model. The main outcome is an estimate for the maximum available emission power (see also the Conclusion Section for a summary). Several details related to the superconducting properties of the emitter are discussed: emitter inductance, plasma frequency, dampening parameter, normal resistance of the intrinsic junctions and resistive shunts, synchronization of the junctions, effect of an external magnetic field, contact resistance and heating.

In Sec. 4 we compute the probability of stimulated emission through a specific rate equation, derived from the Frantz-Nodvik equation, but with three important modifications: (a) A spontaneous emission term, absent in the original equation, is introduced; this is responsible for the start of the emission. (b) Correspondingly, the initial conditions for the solution of the differential equation are different: there is no incoming beam, since the beam is generated inside the active material by spontaneous emission. (c) The population inversion and pumping occur via an external oscillating current, which generates a voltage on the intrinsic junctions, as computed in Sec. 3.

## 2. ANOMALOUS EMISSION AS A CONSEQUENCE OF THE LOCAL MODIFICATION OF VACUUM ENERGY DENSITY IN SUPERCONDUCTORS

Throughout this Section we use units in which $\hbar=c=1$.

### 2.1. Previous Work in Perturbation Theory

In General Relativity and related models (including modifications of Einstein's theory and quantization attempts) the coupling between matter and the gravitational field is described by the tensor $GT_{\mu\nu}$ in the field equations, or equivalently by an interaction term $GT_{\mu\nu}g^{\mu\nu}$ in the Lagrangian. In general, the field equations can also contain a term $\Lambda g_{\mu\nu}$, corresponding to $\sqrt{g}\Lambda$ in the Lagrangian. This term, traditionally named "cosmological term", describes the coupling of the field with the so-called vacuum energy density. By definition, the vacuum energy density is Lorentz invariant, ie it looks the same for any observer in relative uniform motion. It follows that the energy-momentum tensor of vacuum energy density must have the form $const \cdot g_{\mu\nu}$. In the vacuum energy density are usually included the zero-point energies of the quantized fields (including the gravitational field itself). Further contributions to $\Lambda$ originate from the non-vanishing vacuum expectation values of quantum fields in the presence of spontaneous symmetry breaking.

For instance, for a scalar field with vacuum expectation value $\varphi_0$ and Lagrangian $L(\varphi)$, the cosmological term is $-8\pi GL(\varphi_0)$, because the energy-momentum tensor has the form $T_{\mu\nu}=\partial_\mu\varphi\partial_\nu\varphi-g_{\mu\nu}L$. For the electromagnetic field, the part of $T_{\mu\nu}$ proportional to $g_{\mu\nu}$ is $(B^2-E^2)$. Possible contributions of this kind to the global vacuum energy density are supposed to define a uniform background present in the whole universe. Unless there is an exact cancellation of the various contributions, the curvature of the universe should be very large; but this is not observed, and that is the origin of the well-known "cosmological constant problem".

A *local* contribution to the vacuum energy density can arise when the state of a localized physical system is described by a classical field comparable with the vacuum expectation value of a quantum field. We are interested into cases of this kind occurring in condensed matter physics. In this context, the physical systems properly described by *continuous* classical-like fields (also at microscopic level, not just in a macroscopic-average sense as for fluids) are basically: (1) the electromagnetic field in the low-frequency limit, in states where the photons number uncertainty is much larger than the phase uncertainty; (2) systems with macroscopic quantum coherence, described by "order parameters", like superfluids, superconductors and spin systems.

Suppose that, in one of these systems, the field has a constant value $\varphi_0$ in a bounded region and is zero outside. (Consider for instance a container with superfluid helium of constant density.) We can speak of a contribution of the field to the cosmological constant in this region, equal to -8$\pi GL(\varphi_0)$, if it is a scalar field, or the analogous quantity for other fields.

From the classical point of view, one can correctly object that the description of this situation in terms of a local cosmological constant is purely formal, because the gravitational field present is just that due to the superfluid regarded as an energy-momentum source. Moreover, there is no distinction, still at the classical level, between a truly continuous source, like the superfluid wave function, and an incoherent fluid.

The perspective changes if one takes into account short-scale gravitational quantum fluctuations. Suppose to describe gravity with the covariant perturbation theory in the weak-field approximation. The action contains some parameters, and one of these is the effective $\Lambda$ in the considered region. The $\Lambda$ term in superconductors turns out to be much larger than the cosmological background: one typically has $\Lambda/G=10^6$-$10^8$ J/m$^3$ in superconductors, depending on the type, while the currently accepted value for the cosmological background is of the order of $\Lambda/G=10^{-9}$ J/m$^3$ [11]. This is interesting in principle, because it implies in any case a peculiar dynamical condition. At the classical level, however, such a small mass-energy density is irrelevant. As we said, it should be treated quantum-mechanically, being microscopically uniform throughout the superconductor.

In perturbation theory a negative $\Lambda$ (in our conventions) gives gravitons a small real mass, while a positive $\Lambda$ gives them an imaginary mass, ie creates an instability. This seems to suggest a non-trivial role of the $\Lambda$ term in quantum gravity, especially in situations with positive local $\Lambda$, like the unusual case of local density maxima in superconductors [12]. Any supposed instability, however, takes us outside the validity range of perturbation theory. Furthermore, there is no evidence that anomalous effects only occur in situations with positive local $\Lambda$. More often, there appears to be a local $\Lambda$ oscillating in time (e.g., in layered superconductors with high-frequency supercurrents).

The idea, then, is to look for a more fundamental mechanism. We hypothesized earlier [13] that the $\Lambda$ term can be particularly relevant for field configurations with zero scalar curvature. We found a large novel class of off-shell weak field configurations (gravitational vacuum fluctuations) having this property, and studied their modification under the effect of a $\Lambda$ term. Although the fluctuations themselves are very strong and bear a large virtual mass, we found that the effect of the $\Lambda$ term upon them is small, corresponding only to the mass-energy equivalent of $\Lambda$ itself.

In this approximation there is no amplification, ie the $\Lambda$ term does not cause any appreciable variations of the virtual mass density. In general in quantum mechanics an oscillating charge [mass] source emits photons [gravitons], the emission probability being proportional to the square of the source. Then for gravitons the probability is proportional to $\Lambda^2$ and very small, because it also contains the small coupling $G$ [14]. Note that we are talking here of source in a virtual sense, as intermediate state in a quantum process. Such sources would generate the virtual gravitons we called for as possible explanation of the anomalous effects [9].

### 2.2. Vacuum Fluctuations with Large Virtual Mass, in Strong-Field Regime

In subsequent work [15] we thus extended the concept of "dipolar" virtual mass fluctuations to the strong-field case. The √g volume factor in the gravitational action is relevant in this case, and the fluctuations are not exactly dipolar any more. We obtained a wider set of vacuum field configurations, with large virtual masses, present at any length scale. We also showed that this set has finite functional measure, so that the fluctuation effects indeed enter physical averages.

And yet an "amplification" effect of the Λ term is still missing. If we insert a static Λ into the zero-mode equation

$$\frac{rg_{rr}'}{g_{rr}^2} - \frac{1}{g_{rr}} + 1 = 0,$$
(2.1)

the solution will be of the same form as for zero-curvature, but with a slightly different virtual mass. The mass variations will be again of the order of the space integral of the mass density equivalent of Λ/G. Let us show this in detail. The Λ term appears in the zero-mode equation as a constant source term. We already considered in [15] source terms with spatial oscillations, leading to "excited" zero-modes, for instance with the sources sin(*ns*). In that case, the explicit solution of the zero-mode equation was only approximate, although we knew that the exact solution satisfies the zero-mode condition exactly. Here, since the Λ term in the Lagrangian includes a factor √g, this factor is eliminated from the equation and we obtain an explicit exact solution. The Λ term is first supposed constant in time.

Apart from factors 4π, writing $A=g_{rr}$, $B=g_{00}$, the zero-mode equation becomes

$$\sqrt{|AB|}\left(\frac{rA'}{A^2} + 1 - \frac{1}{A}\right) = \sqrt{g}\,\frac{\Lambda}{G}$$
(2.2)

but $\sqrt{g} = \sqrt{|AB|}\,r^2$, therefore the equation is simplified and the solution is

$$\alpha(s) = \frac{1}{s}\left[\int\left(1 + t^2\tilde{\Lambda}\right)dt + k\right]$$
(2.3)

where $\alpha=1/A$, $k$ is an integration constant and $\tilde{\Lambda}$ is the adimensional value of $\Lambda$ after rescaling to the size $r_{ext}$ of the fluctuation: $\tilde{\Lambda} = r_{ext}^2 \Lambda$. We obtain, denoting by $r_0$ the size of the region where $\Lambda$ is not zero and by $s_0=r_0/r_{ext}$ the corresponding adimensional quantity:

$$\alpha(s) = 1 + \frac{k}{s} + \frac{1}{s}\frac{\tilde{\Lambda}s_0^3}{3} \tag{2.4}$$

By correspondence, $k$ must be the virtual adimensional mass $\tilde{M}$ in the absence of $\Lambda$. The factor $\frac{\tilde{\Lambda}s_0^3}{3}$ gives the additional mass due to the cosmological term. Remembering that adimensional masses are scaled according to the rule $M = \frac{\tilde{M}r_{ext}}{2G}$, we find that the additional mass is of the order of $M_\Lambda = \frac{\Lambda r_0^3}{G}$. This is the mass resulting from the energy-density $\Lambda/G$ integrated over the volume $r_0^3$ where $\Lambda$ is not zero.

## 2.3. A Possible Connection between an Oscillating Local Λ-Term and Virtual Mass Fluctuations

In this Section we offer a new conjecture which has a definite intuitive justification but still lacks a proof. Namely, we try to show that an oscillating $\Lambda$ term generates synchronous oscillations in the virtual mass density.

We have seen that strong vacuum fluctuations with large virtual mass exist in quantum gravity. We would like now to display a connection between an oscillating $\Lambda$ and virtual mass fluctuations. The main idea is, that a local time-variable $\Lambda$ term causes changes in the fluctuations spectrum. These changes have observable physical effects. Like in the Casimir effect, there exists a definite interface between the real and the virtual world. For the Casimir effect, the interface are the metal plates which cut the virtual electromagnetic modes. Here $\Lambda$ is not a static cut-off, but oscillates in the superconductor with a certain frequency. $\Lambda$ appears in the "state" equation of the vacuum fluctuations as an external source; when it is oscillating, it could be regarded as a forcing term. We know that $\Lambda$ is very small and we have seen in the previous section that its static effects are minimal, namely tiny shifts in the virtual masses. If we model vacuum fluctuations by a collection of harmonic oscillators, then a static $\Lambda$ term would shift slightly their equilibrium positions. Even as a forcing term, $\Lambda$ would be inefficient, causing small oscillation amplitudes proportional only to $\Lambda$ itself.

In gravitation the vacuum fluctuations are much stronger than in QED, and virtual Casimir-like forces can be much stronger, too. We think that the $\Lambda$ term could "attract" to its own frequency some of the natural virtual amplitude present at nearby frequencies. This change of spectrum would be observable, and its amplitude proportional both to $\Lambda$ and to the pre-existing natural virtual amplitude.

This frequency change of an oscillator in response to an external, possibly weak "pilot" signal is a phenomenon also known as "entrainment" of the oscillator [16]. When many weakly interacting oscillators with different proper frequencies are involved, one often speaks of "synchronization" of the oscillators. It is a phenomenon widespread in nature and well described by general mathematical models like for instance the Kuramoto model [31]. A large ensemble of coupled oscillators can be in synchronized or non-synchronized phases, depending on the values of the coupling parameter, on the natural frequency spread and on the amplitude of the pilot signal. Our dipolar vacuum fluctuations are not exactly an ensemble of oscillators, but a definite analogy can be drawn.

In the functional integral all these fluctuations have the same probability, because their action is zero. (A slightly different weight for the configurations could originate from the functional measure, which is however unknown.) We model the evolution of the fluctuations as a random process, like in a Montecarlo simulation: the system makes frequent attempts at transitions to different states, and the transition probability is given by $\exp(-\beta\delta m)$, where $\beta$ is the analogue of an inverse temperature $\beta=1/kT$ and $\delta m=|m-m'|$ is the mass difference between the initial and final state. Transitions are more likely when the mass difference is small, because in that case the functional change in the configuration is also small.

Transitions between fluctuations of different mass effectively appear as mass oscillations. Let us consider a set of fluctuations with mass distributed within a certain interval $\sigma_m$ about a reference mass $M$ which can be very large, up to $M \approx 10^{57}$ cm$^{-1}$ $\approx 10^{22}$ g at the "condensed matter" scale $r_0=10^{-9}$ m, ie the scale of the local $\Lambda$ [15].

Let the mass take discrete values $m_i$. With a large attempt frequency $f_0$, the system makes attempts at transitions, with probability $\exp(-\beta\delta m)$. The result is a temporal sequence of mass values $m_i(t_j)$. By Fourier-analyzing this sequence, we should obtain a spectrum with a broad maximum around some multiple $n$ of the attempt period $T_0=1/f_0$. (Of course, the parameters $f_0$, $\beta$ and $\delta m$ all have to be normalized at the same time.) This means that the average transition time is $\approx nT_0$. The average is made with transitions with different $\delta m$ and different probabilities; the dominant transitions are those with smaller $\delta m$.

We know that a static $\Lambda$ term changes the masses $m_i$ by an amount $M_\Lambda=\Lambda V/G,$ because it enters the "state equation" of the fluctuations as a source term (eq. 2.2). Call $f_\Lambda$ the oscillation frequency of a varying $\Lambda(t)$ (with $f_\Lambda<f_0$). Can we expect that the transition probability is affected, and thus the spectrum of $m_i(t_j)$?

The mass transitions are not strictly periodic, while the external signal $\Lambda(t)$ is, so we expect that in some case the transition will be easier, when $\Lambda(t)$ decreases $\delta m$, or viceversa. Since the amplitudes of the mass oscillations are much larger than the amplitude of $M_\Lambda$, we expect that the variations in the spectral amplitudes of the fluctuations are proportional to $M_\Lambda$ but also to the initial amplitude of the fluctuations. This is the analogue of the frequency entrainment of an oscillator by an external pilot signal, except that the fluctuations are not exactly like an oscillator; they are random transitions without a sharp proper frequency, and so we speak of a change in their spectrum, instead of entrainment of their frequency.

A typical feature of phenomena of frequency entrainment is, that the amplitude of the entrained signal is inversely proportional to the difference between the pilot signal and the proper frequency of the oscillation [17]. We can check that this is also the case in our simple model of transitions. If we consider a transition with probability close to 1, which means that the inverse temperature parameter β is small (ie, the temperature large, and transitions quite probable), we can write the probability as $P \approx 1 - \beta \delta m + (\beta \delta m)^2/2$. If $\delta m$ changes by an amount $M_\Lambda$, the transition can be favoured or hindered and the average probability change $\Delta P$ is of order $(\beta \delta m)^2$. The change in the average transition frequency is $\Delta f_{av} \approx \Delta P / T_0$ and so finally $\Delta f_{av}/f_0 \approx (\beta M_\Lambda)^2$.

Since in the current approximation the transition probability is of order 1, then $\beta M_\Lambda$ is much smaller than 1, and of the order of $M_\Lambda/\delta m$. Therefore $\Delta f_{av}/f_0 \approx (M_\Lambda/\delta m)^2$. This gives a relation between the frequency shift, or entrainment of the fluctuations, and the fluctuation amplitude which is shifted in frequency. We are interested into frequency shifts of fluctuations with an amplitude $\delta m$ which is much larger than $M_\Lambda$. The squared inverse proportionality between the frequency shift and the amplitude sets a limit on the deformation of the fluctuations spectrum.

In order to assess the final effect of the Λ term, and judge which deformations of the spectrum are observable, we still need to know how the "natural" spectrum is. Presently we do not have any information about the shape of this spectrum, but just about the single modes. A further missing piece of information is the number of fluctuations or mass oscillations present in a given region of space. We can not know this until we know the upper cut-off on their mass or some other UV cut-off. The total change of amplitude in mass fluctuations depends on the number of field oscillators; the amplitudes displaced from their original frequency add coherently (because they all follow the oscillations of the pilot Λ) and therefore this sum process between modes is crucial.

Also in systems of weakly interacting classical oscillators (and our fluctuations are certainly weakly interacting) the emergence of phenomena of synchronization or collective entrainment by an external pilot signal depends on the presence of a large number of oscillators. We shall see an example of this behaviour in the Josephson junctions arrays: their synchronization is observed only above a certain number of junctions.

### 2.4. How to Define the *A* and *B* Coefficients for Virtual Graviton Emission?

The last logical step required is: how to relate the occurrence of virtual mass oscillations at the same frequency of $\Lambda(t)$ (but with larger amplitude) to the emission of virtual gravitons with Einstein coefficients *A* and *B* of spontaneous and stimulated emission? Such a relation has been previously demonstrated by Rogovin and Scully [27] for the electromagnetic emission in a Josephson junction under finite voltage. In that case, the macroscopic classical picture of the oscillating dipole is well complemented by the quantum-mechanical picture of electrons in a collective wave function, undergoing quantum tunnelling between two states. In the case of the Josephson junction, the oscillating charge/current is that of the Cooper pairs. In the gravitational case, the oscillating mass is not directly the mass of the Cooper pairs, it is $M_{\Lambda(t)} = \Lambda(t)V/G$, amplified by "entrained" vacuum fluctuations. $\Lambda(t)$ has a definite expression in terms of the pairs density ρ [12]:

$$\Lambda = -\frac{1}{2m}\left[\hbar^2(\nabla\rho)^2 + \hbar^2\rho\nabla^2\rho - m\beta\rho^4\right], \tag{2.5}$$

where β is the second Ginzburg-Landau coefficient and *m* is the Cooper pair mass.

## 3. CERAMIC SUPERCONDUCTING "EMITTERS" MODELLED AS SERIES OF INTRINSIC JOSEPHSON JUNCTIONS

We are currently performing precise measurements of the behaviour of melt-textured ceramic superconductors under the conditions of Podkletnov's latest experiment [9]. The superconductors are subjected to powerful high-frequency current or current pulses. Since the superconductors are expected to emit Josephson radiation under these conditions, and possibly (according to [9]) also anomalous gravitational-like radiation, we shall call them "emitters".

Our emitters are made of melt-textured YBCO and are reduced in size (diameter 5 cm). They can be modelled as a stack of intrinsic Josephson junctions, in agreement with all modern studies of the conduction of cuprates in the *c* direction [18]. Alternatively, one can describe the material as a homogeneous superconductor with complex conductivity $\sigma=\sigma_1+i\sigma_2$; the conclusions are compatible with the Josephson junctions model, but the information available in the literature about $\sigma_1$ and $\sigma_2$ in the cuprates in dependence on T and other parameters is scarce (one should extrapolate from Bardeen theory [19]). The intrinsic Josephson junctions model is furthermore better suited to describe the electromagnetic emission of the material, which is important because partly related to the anomalous emission.

The appropriate parameters of the intrinsic Josephson junctions will be discussed in detail below. For now we observe that, as confirmed by numerical simulations, being the emitter inductance and capacitance $L_E$ and $C_E$ much smaller than those of the external circuit, they do not substantially affect the oscillation frequency. The simulations also show, as can be expected since the material "must" anyway conduct with excellent values of σ at MHz frequency, that the normal current in the junctions adjusts itself to a value $I_n$ ($<<I_s$; $I_s$ supercurrent) such that the voltage-per-plane corresponds to an AC Josephson frequency equal to the external frequency. This AC Josephson frequency is in turn the same of the Cooper pairs interplane tunnelling, and so the same of the anomalous emission.

### 3.1. Emitter Inductance, Capacitance, Plasma Frequency, Dampening Parameter

Let us find the inductance of the emitter as a series of Josephson junctions. There are ≈$10^7$ junctions, since the emitter thickness is about 1 cm and the inter-plane spacing about 1 nm. The inductance of a single junction is $L \approx \phi_0/I_J$, with $\phi_0=h/2e\approx2\cdot10^{-15}$ Wb and $I_J$ critical current (at least $10^4$ A in our case). We find $L \approx 10^{-19}$ H. With $10^7$ crystal planes, the total inductance is: $L_E = 10^{-12}$ H, to be compared with $L_L\approx10^{-6}$ H of the external circuit. Each layer is seen simply as a very wide junction in this model; in fact there will be distinct coherence regions in the layer, each in parallel with the others, but the final result is the same.

Next we find the capacitance of the emitter as a series of junctions. For a couple of crystal planes (cross-section $S \approx 20$ cm$^2$, distance $d \approx 1$ nm, relative dielectric constant $\varepsilon$ of the order of 10), we have $C = \varepsilon_0 \varepsilon S/d \approx 10^{-4}$ F. Dividing by $10^7$, the total capacitance is found to be $C_E \approx 10^{-11}$ F, to be compared with $C_L \approx 10^{-8}$ F of the external circuit.

Therefore the proper frequency of the Josephson junctions, also called plasma frequency $f_P$, is $f_P = 1/2\pi\sqrt{(L_L C_L)} \approx 10^{11}$ Hz. It is natural to expect that this proper oscillation does not influence the behaviour of the system under the effect of an external forcing frequency typically smaller, of the order of 0.1-10 MHz. Note that $f_P$ is the same for the emitter and for any single junction, because the capacitances and inductances in series scale in the opposite way. The formula for $f_P$ can be easily re-written as follows, with reference to a single junction [18]:

$$f_P = \left(\frac{I_J}{2\pi\phi_0 C}\right)^{1/2} = \left(\frac{j_J d}{2\pi\phi_0 \varepsilon \varepsilon_0}\right)^{1/2} \tag{3.1}$$

The McCumber parameter of a junction $\beta_c$ is defined by $\sqrt{\beta_c} = 2\pi f_P RC$. This is connected to the hysteresis of the I-V curve of the junction, because $\sqrt{\beta_c} = (4/\pi)I_J/I_r$, where $I_r$ is the so-called return current. For $\beta_c < 1$ we have over-dampened, non-hysteretic junctions; for $\beta_c > 1$ we have under-dampened, hysteretic junctions. With the data above, one finds for the single junction $\sqrt{\beta_c} \approx 10^8 R$. Therefore our junctions are strongly over-dampened, because $R$ is less than $10^{-10}$ $\Omega$ for the single junction.

## 3.2. Normal Resistance of the Intrinsic Josephson Junctions in YBCO and "Resistive Shunts"

The intrinsic Josephson effect has been observed in YBCO as clearly as in BSCCO. Kawae et al. [20] give evidence of I-V curves in YBCO with multiple branches and hysteresis, very similar to those reported by Kleiner and Muller for BSCCO [18]. This can only be seen, however, in very small samples, with area about 0.25 μm$^2$. In larger samples, grain borders or other defects act as low-resistance shunts. The total resistance is essentially determined by these shunts, and so depends on the micro-structure and not just on the material. The junctions become non-hysteretic, because the McCumber parameter $\beta_c$ is proportional to $R$, and a small $\beta_c$ means no hysteresis. For this reason, the presence of the single junctions can not be seen in the I-V curves of "large" samples. From the practical point of view, all this does not disturb much, except that it is impossible to know in advance the resistance of our material, it depends on the micro-structure. The CRC data [25] is only an indication: $\rho = 5 \cdot 10^{-5}$ $\Omega$m at room temperature, implying $R_E = 10^{-4}$ $\Omega$ for our emitter. According to CRC, there is only a small variation in the normal resistance between room temperature and 100 K.

Ref.s [20] and [21] also allow to estimate the normal resistance of the employed samples. For a stack of 80 junctions, ref. [20] gives at 4.2 K a critical current of 0.1 mA (40 kA/cm$^2$);

the return voltage is 0.2 V and the slope of the I-V characteristic in the single normal branch is 800-1000 Ω. The $I_cR$ product ($I_c$ critical current of YBCO material) is therefore 2-3 mV per junction; the material has $T_c$=43 K, so the BCS prediction is about 10 mV. The resistivity computed from the data above is $3 \cdot 10^{-3}$ Ωm. In ref. [21] the junctions have size 0.65×0.85 mm$^2$ and $R$=2 μΩ per crystal plane at the peak resistance value (84K; $T_c$=93 K). This gives ρ=$1.1 \cdot 10^{-3}$ Ωm.

The measured resistance of our emitter, including contacts, is $3 \cdot 10^{-4}$ Ω at room temperature and $5 \cdot 10^{-6}$ - $12 \cdot 10^{-6}$ Ω at 77 K. For noble metals the resistivity varies by a factor 5-10 between 300 K and 80 K; for iron more than 10 times. Alloys with noble metals show smaller variations, typically a factor 2-3. At 77 K the DC resistance of YBCO alone vanishes, therefore the residual resistance is that of the metal layer and of the contact. On the contrary, the $3 \cdot 10^{-4}$ Ω at room temperature are essentially due to the YBCO, in agreement with the CRC data above.

The aim of works like [20] is to see the features of the microscopic junctions (their resistance, capacitance, impedance, McCumber parameter), in view of possible applications for fast electronics, microwave collective emission or detection etcetera. In our case we will be satisfied to know that the intrinsic Josephson junctions are active and syncronized. Seeing their individual signature is not essential, actually impossible in a bulk sample. We believe they have to be syncronized, because otherwise they could not sustain the oscillating current; they would instead pass into a purely resistive state and the oscillating current would be normal current. The voltage on the emitter would then be of the order of $10^4$ A $\cdot 10^{-4}$ Ω ≈ 1 V.

### 3.3. First Estimate of the Total Anomalous Radiation Energy $U_{max}$ and of the Dissipation in the Emitter

We have found a simple way to make an order-of-magnitude estimate of $U_{max}$, which agrees with the results of the detailed simulations (see below). Assume that the voltage-per-plane in the emitter is given by the Josephson relation $V=hf/2e=\phi_0 f$. Here $f$ is the frequency of the *external* circuit. This is necessary in order that the external current flows in the emitter as supercurrent (except for the small normal component $I_n$ which gives the finite voltage). The numerical simulations confirm this coincidence of external frequency and Josephson frequency.

Take, for instance, an external circuit with frequency $f$=10 MHz, current 1 kA and dampening time τ=$10^{-4}$ s. The voltage over a single junction is $V = 2 \cdot 10^{-15} f = 2 \cdot 10^{-8}$ V. The total number of junctions in 1 cm thickness is ≈ $10^7$ (each is 1.17 nm). Thus the total voltage on the emitter is 0.2 V. The IV product in the emitter, also called DC-power $P_{DC}$ is $P_{DC}=IV$=200 W. In the time τ this makes available in the emitter an energy of the order of 20 mJ.

The electromagnetic emission generated in the AC Josephson effect has an energetic efficiency which is typically of the order of 10% [22]. We suppose that the anomalous emission is associated with the electromagnetic emission, ie a graviton is emitted together with the photon at each Cooper pair tunnelling, at the same frequency and with an emission probability of the same magnitude order. So the total energy $U_{max}$ of the anomalous radiation is ≈ 2 mJ.

Approximately the 80% of the IV product is wasted for the emission and dissipated as heat. This does not cause any serious temperature increase. A reasonable estimate for the thermal capacity of our emitter is 200 mJ/K; the temperature increase of the bulk is therefore negligible. This thermal capacity can be obtained as follows, using, for instance, data from Tilley [23] for the specific heat of YBCO as a function of temperature. The density of YBCO depends on the cell parameters, which are variable. Taking for instance $a$=0.38 nm, $b$=0.39, $c$=1.17 nm (Waldram [24], $YBCO_{7-x}$, $x$=0.4), one finds a unit cell volume of $1.7 \cdot 10^{-28}$ $m^3$. The unit cell has a total mass of 560 a.m.u. This gives a density of 6300 $kg/m^3$ (measured value: about 6000 $kg/m^3$). Therefore the thermal capacity of the emitter, supposed it has volume 20 $cm^3$, is 200 mJ/K. 10 $mJ/Kcm^3$ corresponds to 1.5 mJ/Kg.

Note that the IV product on the emitter does not depend on the external load resistance $R_L$. If we can reduce $R_L$, we will increase τ and so proportionally increase the target energy. There is a practical problem with a small $R_L$, however: increased dissipation at the S/N contacts and possible damage to the external capacitors (see below).

The normal current $I_n$ in the emitter can be computed through the relation $V=I_n R_E$. The resistance $R_E$ can roughly be guessed from the data in the literature. For instance, if $R_E \approx 10^{-4}$ Ω (CRC) one finds $I_n \approx 10^2$ A; if $R_E \approx 10^{-3}$ Ω [21], $I_n \approx 10$ A. The pure ohmic heating is therefore irrelevant. The figure for $I_n$ gives the *total* normal current. The density of normal current varies locally, as more current flows in the shunts with smaller resistance.

## 3.4. Synchronization of the Emission. Magnetic Field. Literature Review

Emission of laser-like, coherent radiation from intrinsic Josephson junctions has been observed by Barbara et al. [26] when the junctions are enclosed in a microwave cavity, which serves to impose a definite common oscillation frequency to the junctions. In our case the common frequency is set by the external circuit. The superconductor just follows the external oscillation. The general response of a superconductor to an AC voltage in the KHz-MHz range is to exhibit a small impedance, with small resistive and inductive components (related to the $σ_1$ and $σ_2$ mentioned above). For cuprates this is still true, independently from their intrinsic-Josephson structure. We are assuming that in large samples with resistive shunts the intrinsic Josephson structure, un-observable in the I-V curves, is nevertheless active for coherent electromagnetic and anomalous emission.

In the cited works, the array size is comparable or larger than the free-space radiation wavelength λ=2 mm. The emission frequency corresponds to a high-Q resonance in the structure formed by the array and the resonator ground plane. The power coupled to the detector is actually transmitted through a non-linear transmission line, so λ is not exactly that of free space. The detector is itself made of junctions, and is very close.

At MHz frequency, λ is clearly much larger than the system's size. In our case, λ is comparable to the system size, but the anomalous radiation is supposed to be only virtual (compare also Sect. 4).

The transition of the junctions array to a coherent state was predicted by Bonifacio et al. [29] on the basis of the formal analogy between Josephson junctions arrays and free electron lasers. Earlier, such a quantum coupling mechanism was predicted by Tilley [30]. Jain [22], on the contrary, describes classical synchronization of over-dampened Josephson junctions with resistive shunts and low efficiency, about 1%. Note that our junctions are over-dampened because $\omega_c RC \ll 1$, due to resistive shunts and low resistance. Our junctions are driven from an external AC current, however, so they keep oscillating in spite of being over-dampened.

An extensive literature search and study about the onset of synchronization in arrays of artificial and intrinsic Josephson junctions confirmed that an external load causes synchronization. In many experiments and simulations, the external load is just an RLC circuit as in our case. Our simulations with few junctions clearly exhibit synchronization. Note that in our case, not only has the external circuit a definite proper frequency, but the initial conditions are such that the circuit oscillates from the beginning, while in several other experiments and simulations the junctions are DC biased and coupled to a resonant circuit which is initially passive.

In general, a magnetic field in the *ab* direction should increase the inter-plane coupling (Kleiner et al., [18]). It does so, however, at the price of decreasing the critical current $I_J$. Other coupling mechanisms should be more effective in our case, in particular the external driving frequency and the normal current ("quasi-particles" current). The numerical simulations predict that, paradoxically, a lower $I_J$ (provided still larger than the external current $I_0$) increases the emitter voltage and thus the DC Josephson power. Therefore it can be helpful to apply an uniform magnetic field to the emitter. The dependence of $I_J$ on the field is not strong and $V_E$ is inversely proportional to $I_J$, so the power gain is not expected to be dramatic. Also, it is difficult to define the degree of uniformity needed.

In the computations by Rogovin and Scully [27] the magnetic field appears explicitly. Sometimes it couples e.m. normal modes with different polarizations (see also the work by Almaas and Stroud below). In principle, certain modes of the Josephson junctions would be decoupled from the radiation in the absence of a static magnetic field.

Acebron et al. [31] and K. Wiesenfeld et al. [32] consider several synchronization problems, among them the one we are concerned with. They find that Josephson-junction arrays connected in series through a load exhibit "all-to-all" (that is, global) coupling. A schematic circuit is given, with ideal junctions in series coupled through a resistance-inductance-capacitance load; in parallel to both is a bias current generator.

A model for a large number of Josephson junctions coupled to a cavity and an attempt at an explanation of the experiment by Barbara et al. for 2D arrays was given in [33]. The synchronization behavior was reproduced. Junctions are under-dampened, with non-zero C. A bias current is taken such that each junction is in the hysteretic regime. Depending on the intial conditions, the junctions may work in each of two possible states, with zero or non-zero voltage. In the latter case, the phases vary with time and the junctions are called "active".

Filatrella and Pedersen [34] find that the transition from a state where the junctions are essentially oscillating at the unperturbed frequencies to one where they oscillate at the same frequency occurs above a threshold number of active junctions, in agreement with the experimental results by Barbara et al. A subsequent work [35] studied conditions when there is no threshold.

In general, the model employed in the papers above includes a global, "classical" coupling (external oscillating circuit), while Barbara et al. in their PRL article stressed the fact that the coupling is local and typically quantum-mechanical, with stimulated emission. For this reason, probably, is the threshold behaviour not properly reproduced.

In our simulations we feed an oscillating current into the junctions, instead of a DC bias. The authors of [35] suppose that all junctions have the same $R$ (for us, not necessarily, because variations are compensated by variations in the normal current $I_n$), and let instead the critical current $I_J$ vary (our simulation also supports this; the voltage on the junctions depends on $I_c$; in any case we have $I<I_J$).

Almaas and Stroud [37] give a theory of 2D Josephson arrays in a resonant cavity. They consider the dynamics of a 2D array of under-dampened junctions placed in a single-mode resonant cavity, in the limit of many photons. The numerical results show many features similar to the experiment by Barbara et al., namely: (1) self-induced resonant steps; (2) a threshold number of active rows; (3) a time-averaged cavity energy which is quadratic in the number of the active junctions. They predict a strong polarization effect: if the cavity mode is polarised perpendicular to the direction of current injection in a square array, then it does not couple to the array and no power is radiated into the cavity. In the presence of an applied magnetic field, however, a mode with this polarisation would couple to an applied current.

### 3.5. Contact Resistance and Heating

In experiments on intrinsic Josephson junctions there is usually a transport current along the c axis, fed in from a generator through special contacts on the top and bottom of the samples. At the contacts, most of the external current is converted into super-current. The same should happen in our case, but our current is large and there is the problem of contacts over-heating. (The current is well below $I_c$, but this is so because melt-textured materials have especially large $I_c$.) If the material is driven normal near the contacts, all the mechanism of superconducting conduction and Josephson tunnelling is lost. The material can be driven normal also because contact is not uniform and local current density exceeds $J_c$.

We have seen that dissipation and heating in the bulk of the emitter can be disregarded. We must then check heating at the contacts. In work by Takeya et al. [18] the heat diffusion in BSCCO is taken into account. A heat diffusion length $l$ can be defined, both in the ab and c directions. Heat delivered at one point spreads over a volume of approximate size $l^2_{ab} \cdot l_c$. Knowing the specific heat of the material, one can compute the temperature increase of that volume. We are interested only in $l_c$, since heat is generated at planar contacts.

Next we need a guess for the surface resistance of the contacts $R_c$. Take for instance $R_c=10^{-5}$ Ω (IoP Handbook [19], Sect. B.5), ie $\rho_c=2\cdot10^{-4}$ Ωcm$^2$; then for larger or smaller values all scales in proportion. Heat generated at the contacts is of the order of the total energy ($10^2$ J) multiplied by the ratio $R_c/R_{load}$. For instance, with $R_{load}=0.1$ Ω we find a dissipation of $10^{-2}$ J.

The heat diffusion length is given by the formula $l=2\sqrt{(Kt/c)}$, where $K$ is the heat conductivity, $c$ the specific heat, t the duration of the pulse. For BSCCO, Takeya et al. give $K=0.25$ W/mK, $c=2$ kJ/Km$^3$. For YBCO, $K=15$ W/mK along ab ([24], p. 254), $c=10$ kJ/Km$^3$ (see above). For instance, with a pulse duration t=0.5 μs, we find $l$=1.7 mm and the interested volume is $3\cdot10^{-6}$ m$^3$; its thermal capacity is 30 mJ/K. The thermal capacity of a copper feeding electrode, however, is bigger, so it takes much of the heat. (One should also take into account a possible indium layer between YBCO and copper.) The temperature increase would then be negligible.

So, supposed a surface resistivity of the order of $10^{-4}$ $\Omega cm^2$ can be obtained, there would be room for a reduction of the external resistance $R_{load}$, admitted this is possible in practice. If $R_{load}$ is smaller, then the dampening time $\tau$ is larger and the total energy $U_{max}$ of the anomalous radiation increases in proportion. Dissipation in the bulk of the emitter and in the contacts also increases in proportion to $\tau$.

## 3.6. Simulation of a Josephson Junction Inserted in a RLC Circuit, in the RSJ Model

According to the RSJ model (resistively-shunted junction), a Josephson junction can be represented as a non-linear circuit element obeying the Josephson effect equations below, plus an ohmic resistance $R$ in parallel. In the purely Josephson element flows only supercurrent while in the resistance flows normal current. We have seen that when the Josephson junction is placed in an external oscillating circuit with large $C$ and $L$, it should not influence the external current. This is confirmed by the simulation below and is true also for many junctions in series. Therefore we first simulate one single junction and then we shall consider the synchronization of several junctions.

The two fundamental equations of the Josephson effect are

$$I_s = I_J \sin\phi, \qquad (3.2)$$

where $I_s$ is the supercurrent in the junction, $I_J$ is the critical current and $\phi$ the phase difference over the link, and

$$\phi' = \frac{2e}{\hbar} V, \qquad (3.3)$$

where the prime denotes time derivative and $V$ is the voltage applied to the junction. According to the RSJ model, $V = RI_n$, where $I_n$ is the normal current flowing in the normal resistance $R$ of the junction, parallel to $I_s$.

Only $I_n$ generates a voltage in the emitter, but both $I_n$ and $I_s$ flow in the external resistance and inductance ($I_s$ after conversion to normal) and discharge the capacitor.

Denote $a = \frac{2e}{\hbar} R$ and rewrite (3.3) and the second derivative of (3.2) as follows

$$\begin{cases} \phi' = aI_n \\ I_s'' = aI_J(I_n'\cos\phi - aI_n^2 \sin\phi) \end{cases}. \qquad (3.4)$$

These are the first two equations of a system, whose unknowns are the functions of time $\phi(t)$, $I_s(t)$, $I_n(t)$.

Write the derivative of the Kirchoff equation over the loop including the external load ($L_L$, $C_L$, $R_L$) and the junction

$$\frac{1}{C_L}(I_s + I_n) + L_L(I_s'' + I_n'') + R_L(I_s' + I_n') + RI_n' = 0. \quad (3.5)$$

This is going to be the third equation of the system. Divide by $L_L$ and note that the proper frequency of the external circuit is $\omega = 1/\sqrt{L_L C_L}$. Disregard the last term because $R$ is about $10^{10}$ times smaller than $R_L$. Replace $I_s''$ with the second equation in (3.4), where $I_J$ is denoted $g$. Finally define $b = R_L/L_L$. We find

$$\omega^2(I_s + I_n) + ag(I_n' \cos\phi - aI_n^2 \sin\phi) + I_n'' + b(I_s' + I_n') = 0. \quad (3.6)$$

Isolating $I_n''$, we obtain the final complete non-linear system, where the currents are denoted simply by $s$ and $n$:

$$\begin{cases} \phi' = an \\ s'' = ag(n'\cos\phi - an^2 \sin\phi) \\ n'' = -\omega^2(s+n) - ag(n'\cos\phi - an^2 \sin\phi) - b(s'+n') \end{cases} \quad (3.7)$$

Summarizing, the symbols and typical magnitude orders of the parameters are, in SI units,

$$\begin{aligned} &n = I_n(t) \\ &s = I_s(t) \\ &a = \frac{2e}{\hbar} R = 3 \cdot 10^4 \\ &b = \frac{R_L}{L_L} = 1.5 \cdot 10^4 \end{aligned} \quad (3.8a)$$

$$\begin{aligned} &R = R_E / N = 10^{-11} \\ &\omega = 3.3 \cdot 10^6 \\ &g = I_J = 5 \cdot 10^4 \end{aligned}$$
$$(3.8b)$$

The initial conditions at time $t=0$ (when the external circuit is closed) are the following:

$$\phi(0) = 0$$
$$I_n(0) = 0$$
$$I_s(0) = 0 \qquad (3.9)$$
$$I_s{'}(0) = 0$$
$$I_n{'}(0) = I_0 \omega$$

At the time $t=0$ the external circuit begins to oscillate, starting from a state in which the capacitor is fully loaded. The initial value for $I'_n(0)$ is standard for an RLC circuit. $I_0$ is the maximum external current, which depends on $V$, $C_L$ and $L_L$ as

$$I_0 \approx V \sqrt{\frac{C_L}{L_L}} \qquad (3.10)$$

Note that $I_s$' is initially zero due to eq. (3.2) and (3.3), since $V$ is initially zero. It is interesting to note that in spite of this, $I_s$ rapidly grows and becomes almost equal to $I_0$ in the emitter, where $I_n$ stays small (see below).

With these initial conditions the equation system (3.7) can be solved numerically through the Runge-Kutta method. The result is clear: for $I_0<I_J$ (which is usually the case) all functions oscillate with the external frequency. With the parameters above, $\phi$ oscillates between 0.1 and –0.1, while $I_n \approx 10$ A. For $I_0=I_J$, the phase makes a complete oscillation in the period of the circuit oscillation.

A refined version of the RSJ model includes a junction capacitance $C_J$ in parallel to the resistance [28]. For high frequency, the capacitive channel can become important. We have seen that $C_J \approx 10^{-4}$ F, so the impedance of the C-channel at $\omega \approx 1$ MHz is of the order of $10^{-2}$ Ω, much larger than $R_E \approx 10^{-11}$ Ω. It should therefore be legitimate to disregard $C_J$. For a check, we included a capacitive channel in the numerical simulation. The results are at first sight puzzling, because in this case the capacitance of the Josephson junction affects the circuit behaviour much more than the (smaller) external capacitance $C_L$; but this is an artefact, because eventually we want to simulate a large number of junctions in series, and in that case their total capacitance will be small, so $C_L$ will actually dominate and the C-channels of the junction carry very little current.

Let us write the equation for 2 junctions in series:

$$s_1 = g \sin \phi_1$$
$$s_2 = g \sin \phi_2$$
$$\phi_1' = a n_1$$
$$\phi_2' = a n_2 \quad\quad\quad (3.11)$$
$$s_1'' = ac(n_1' \cos \phi_1 - a n_1^2 \sin \phi_1)$$
$$s_2'' = ac(n_2' \cos \phi_2 - a n_2^2 \sin \phi_2)$$
$$\frac{1}{C_L}(n_1 + s_1) + L_L(n_1'' + s_1'') + R_L(n_1' + s_1') = 0$$

In the last equation one isolates $n_1''$ and replaces $s_1''$. In order to find $n_2''$, we note that $n_1+s_1=n_2+s_2$ → $n_1''+s_1''=n_2''+s_2''$ → $n_2''=…$ We so have 6 equations with unknown $\phi_1$, $s_1$, $n_1$, $f_2$, $s_2$, $n_2$. The initial condition is the same, as is easily obtained differentiating the equation $n_1+s_1=n_2+s_2$.

For 3 junctions: the equations for $\phi_3'$ and $s_3''$ are simple. Then current conservation gives $n_3''=n_1''+s_1''-s_3''$. This also holds for the first derivatives, and for the initial condition, which is just the same. And so on.

In this way we check directly the synchronization, at least for few junctions, and regarding a whole crystal layer (with surface of the order of square centimetres!) as a single junction. Phase, voltage and normal current are synchronized. The synchronization also occurs for higher frequency (larger than 10 MHz).

The simulations allow to compute the emitter voltage $V_E$ by multiplying $I_n$ and $R_E$. This voltage turns out to be smaller (typically 10 times smaller, with the parameters above) than the simple estimate based on the relation $V=(h/2e)f$. This relation holds rigorously for a constant voltage, while in our case we have $V=RI_n$, and $I_n$ oscillates.

In addition, the emitter voltage depends on the critical Josephson current $I_J$, and is larger when $I_J$ is smaller (inversely proportional, see below). This could not have been predicted without the simulations, but the qualitative reason is clear. The supercurrent is fixed (approximately equal to the total current) and $I_s=I_J\sin\phi$. If $I_J$ is larger, then the oscillations of $\phi$ are smaller, and so $\phi'$ is smaller too; and $V$ is proportional to $\phi'$. A possible way to depress $I_J$ is to apply a magnetic field. So the magnetic field may be not needed for synchronization, but improves the $IV$ power and $U_{max}$.

It is not easy to understand intuitively how an oscillating $I_s$ is obtained when the voltage itself oscillates. Mathematically, the point is that $\phi$ does not evolve linearly in time, but oscillates in turn, therefore $I_s$ is not perfectly harmonic while $I_0$ is harmonic, and the difference $I_n=I_0-I_s$ oscillates.

On a short time scale, the simulations show that after $t=0$ the normal current, starting from zero, rises quickly and then begins to oscillate from its maximum. The Josephson junctions are very quick ($10^{-11}$-$10^{-12}$ s) to adapt to the least energy configuration, in which most external current is converted into super-current.

Some simulations were run to look for the dependence of the normal current upon $I_J$. It turns out that there is an inverse proportionality. For instance, with $a_1=a_2=3 \cdot 10^4$ one finds the following values:

| Critical current (kA) | Normal current (A) |
|---|---|
| 20 | 25 |
| 40 | 12.5 |
| 80 | 6.1 |
| 160 | 3.1 |

## 4. STIMULATED EMISSION

We have analysed the behaviour of a superconducting emitter, modelled as a series of Josephson junctions, when it is inserted into an oscillating circuit with proper frequency much smaller than the Josephson plasma frequency. We concluded that the junctions are synchronized with the external circuit and we evaluated the normal- and super-current components $I_n$ and $I_s$. It is interesting to compare the situation with the experiments of Ref.s [26,34]. In that case, the junctions are individually biased with a DC, and are synchronized by a passive external cavity [26] or by a passive external circuit [34,35]. In our case the external current, in the MHz frequency range, serves at the same time as bias and coupling device.

The main question now is: does stimulated emission occur, like in [26]? In the absence of a resonant cavity, this can only occur in a single-pass mode, like in optical or maser amplifiers. Each junction is "pumped" and first emits spontaneously. The emitted photons/gravitons (the model applies to both) propagate and stimulate further emission. A representation of a Josephson current as an ensemble of Cooper pairs tunnellings or as an oscillating macroscopic quantum dipole as been given earlier, as mentioned, by Rogovin and Scully. A rate equation appropriate for a single-pass linear photon amplifier has been given by Frantz and Nodvik [36], and subsequently applied to several cases. It has the form

$$\frac{dn}{dt} + c\frac{dn}{dx} = \sigma c n (N_2 - N_1) \tag{4.1}$$

The Frantz-Nodvik equation has the typical structure of a conservation equation describing the longitudinal propagation of a particles beam with light velocity. The equation takes into account the possibility that particles are absorbed or generated at any point. The l.h.s. represents the net variation of the particles density $n(x,t)$ along the beam.

The equation was originally written for photons, σ being the "resonant photon absorption cross section". In other words, the product $\sigma c$ gives the transition probability per unit time. $N_1(x,t)$ and $N_2(x,t)$ give the density of atoms in the levels 1 and 2. The incoming beam is supposed to be monochromatic.

We would like to consider an incident beam with a frequency spread and to make a connection with the basic equation describing spontaneous and stimulated emission, namely

$$\frac{dN_2}{dt} = -B\rho N_2 - AN_2 \,, \tag{4.2}$$

where ρ is the energy density per volume and frequency:

$$\rho = \frac{dE}{Vdf} = \frac{nhf}{df} \,. \tag{4.3}$$

In our case the bandwith ratio *f*/*df* is fixed by the merit factor *Q* of the external oscillating circuit. Therefore

$$\rho = hQn = const \cdot n \,. \tag{4.4}$$

For our purposes, the FN equation needs to be modified and adapted as follows. (a) A spontaneous emission term is added. This is important at early times, because in our case there is no in-going beam in the initial conditions, but the initial photon density $n(x,0)$ is zero everywhere. (b) In the FN equation, the population difference ($N_2$-$N_1$) is a function of *x* and *t*, albeit one which is eliminated in the final solution. We replace that difference with a constant *N* giving the number of tunnelling processes ("transitions") of Cooper pairs in the intrinsic Josephson junctions per unit time and volume.

The constant *N* depends on the super-current, which together with the emitter voltage $V_E$ defines the maximum pumping power $P \approx V_E I_s$. At each transition, there is a certain probability of spontaneous or stimulated emission of a graviton by each Cooper pair. Competing electromagnetic emission can also occur, but being on-shell it does not have the right wavelength for amplification (see below; in [26] the size of the junctions is much larger, and the emission frequency too, *f*=150 GHz).

In the end, we shall mainly be interested into the saturation condition, when stimulated emission dominates and *n* grows rapidly to the maximum value *n*≈*N* allowed by the pumping. In these conditions, the dominant emission is necessarily longitudinal, because only the longitudinal mode is amplified, while spontaneous emission might be preferentially transverse (like for electromagnetic Josephson emission), also possibly depending on the applied magnetic field. "Longitudinal" means here along the *c* crystal axis, orthogonal to the junctions and parallel to the super-current.

*N* is the number of Josephson transitions per unit time and volume, ie the number of layers per volume in the emitter multiplied by *I*/2*e*. Each layer is seen simply as a "giant" junction in the model, though more realistically there will be in the layer distinct coherence regions, each in parallel with the others. If *S* is the superconductor cross section, δ its thickness and *d* the thickness of a single layer, we have

$$N = \frac{j}{2ed} \tag{4.5}$$

Eq. (4.1) becomes

$$\frac{dn}{dt} + c\frac{dn}{dx} = \gamma nN + AN, \qquad (4.6)$$

where γ is a pure number, which we can express in terms of the other parameters by comparison with (4.2), (4.3) and (4.5). We find the relation

$$\gamma nN = \frac{Bh}{2ed}nQj. \qquad (4.7)$$

The factor (*Bh*/*2ed*) contains only fundamental constants or fixed experimental parameters; the factor *Qj* can be tuned in a certain range. Defining the constants

$$\begin{aligned}\alpha &= \frac{Aj}{2ed} \\ \beta &= \frac{Bh}{2ed}Qj\end{aligned}, \qquad (4.8)$$

with ratio α/β=*A*/(*BhQ*), we obtain the rate equation in final form:

$$\frac{dn}{dt} + c\frac{dn}{dx} = \beta n + \alpha. \qquad (4.9)$$

In order to solve it, we define as usual the auxiliary variables

$$\begin{aligned}\xi &= \frac{x}{c} \\ \rho &= t - \frac{x}{c}\end{aligned}. \qquad (4.10)$$

The equation then becomes

$$\frac{dn}{d\xi} = \beta n + \alpha, \qquad (4.11)$$

with solution

$$n = k(\rho)e^{\beta\xi} - \frac{\alpha}{\beta}. \qquad (4.12)$$

Here $k(\rho)$ is an arbitrary function which we determine returning to the original variables and imposing the initial condition $n(x,0)=0$ for any $x$. The final solution is remarkably independent from $x$:

$$n(x,t) = \frac{\alpha}{\beta}\left(e^{\beta t} - 1\right). \tag{4.13}$$

Note that the emission is supposed to be in the positive $x$ direction (this is implicit in the definition of the variable $\rho$).

We can see here a necessary condition for saturation. When the current oscillates, saturation can only occur if the characteristic time $t_c=1/\beta$ of the exponential growth is smaller than the oscillation time. For a magnitude order estimate, consider for instance $Q=10\text{-}100$, $j=5 \cdot 10^6$ A/m$^2$, $d=1.17$ nm; we find that $t_c < 10^{-7}$ if $B > 10^4$ m$^3$/Js$^2$. This is much smaller than the $B$-coefficient of atomic optical transitions. (In principle, we could estimate $B$ from Podkletnov's data, but the uncertainties on his parameters $j$ and $Q$ are too large.)

The overall amplitude $\alpha/\beta$ depends on the known ratio between the Einstein coefficients $A/B = 8\pi f^3 h/c^3$. We find $\alpha/\beta = (\omega/c)^3/Q$. It is known that $\omega$ enters the $A/B$ ratio because of phase-space considerations based on the formula $p=h/\lambda$ for the photon momentum. Therefore re-inserting $\lambda$ we have $\alpha/\beta = 1/(\lambda^3 Q)$. It follows that the ratio $\alpha/\beta$, and thus $n(x,t)$, is large when $\lambda$ is small, as happens for the virtual anomalous radiation: we know from the experiment [9] that $\lambda f \approx 1$ m/s; then for instance with $f=10^7$ Hz we have $\lambda \approx 10^{-7}$ m. At the same time, this shows that the corresponding real electromagnetic radiation with $f=10^7$ Hz and $\lambda=30$ m is strongly suppressed.

Now taking $Q=10$ we find $\alpha/\beta=10^{20}$. The Cooper pairs density in YBCO is at least $10^{26}$ m$^{-3}$, therefore for saturation the exponential factor must be $10^6$, or the exponent $t/t_c \approx 10$.

## 5. Conclusion

The first necessary condition for a sizeable gravitational emission from superconductors subjected to high-frequency currents is that the standard matter/field coupling must be amplified by some microscopic quantum mechanism. We have pointed out the possible existence and nature of such a dynamical mechanism, but a rigorous proof is not yet available. We call "anomalous" any kind of gravitational emission due to this fundamental anomalous coupling, which can only occur when matter is in a coherent state. According to our model, the coupling is mediated by the vacuum energy density $\Lambda$, and amplified by gravitational vacuum fluctuations. The emitted gravitons can only be off-shell, with $\lambda f \ll c$ and have finite propagation range (but can have spin 0 and 1 components). They are virtual particles, ie they exist only as intermediate states of quantum processes.

The second condition is, that the overall energetic balance must be respected. The emitting transitions are Josephson tunnellings of Cooper pairs in the intrinsic junctions of oriented (melt-textured) anisotropic ceramic superconductors. The pumping occurs via a high-frequency current in the $c$ crystal direction. The current has super- and normal-components, $I_n$ and $I_s$. $I_n$ is much smaller than $I_s$, typically $10^3$ times smaller, and is necessary in order to establish the voltage $V_E = R_n I_n$ which supports the oscillations of $I_s$. The maximum available power is $P = IV_E \approx I_s V_E = I_s I_n R_n$. The effective normal resistance $R_n$ of the material depends on its micro-structure (resistive micro-shunts), but numerical simulations show that $I_n$ adapts to $R_n$, keeping $V_E$ constant; therefore the shunts micro-structure is not critical. The emitter voltage depends on the external frequency and on the critical Josephson current $I_J$.

In our laboratory trials, the estimated maximum pumping power is of the order of $10^2$ W, but the emitter voltage $V_E$ could not yet be measured, due to powerful disturbances generated by the external circuit. An increase in the maximum available power is technically difficult, and in any case there are stringent theoretical limits on the maximum voltage present in the superconducting emitter. Further technical problems which emerged from our preliminary trials and planning concern thermal dissipation in the bulk and at normal-superconducting contacts, and the application of a proper magnetic field.

The third necessary condition is, that the pumping power must be exploited as much as possible, and for this it is crucial that a cascade process of stimulated emission is activated. The cascade can occur in a single passage, because the emitting layers are very numerous ($\approx 10^7$), and is governed by the rate equation (4.9) which takes into account the Einstein coefficients $A$ and $B$, the current density $j$ in the emitter and the merit factor $Q$ of the external circuit. A compromise between these parameters should be found, such that the stimulated emission cascade can fully develop in a high-frequency cycle or in short pulses.

## ACKNOWLEDGMENT

This work was supported by the Göde Wissenschaftsstiftung (Goede Science Foundation).